# Efficient Fourier single-pixel imaging with Gaussian random sampling

Ziheng Qiu,[1] Xinyi Guo,[1] Tian'ao Lu,[1] Pan Qi,[3] Zibang Zhang,[1,2,*] and Jingang Zhong[1,2]

[1] *Department of Optoelectronic Engineering, Jinan University, Guangzhou 510632, China*

[2] *Guangdong Provincial Key Laboratory of Optical Fiber Sensing and Communications, Jinan University, Guangzhou 510632, China*

[3] *Department of Electronics Engineering, Guangdong Communication Polytechnic, Guangzhou 510650, China*

*charles.cheung.zzb@gmail.com

**Abstract**

Fourier single-pixel imaging (FSI) is a branch of single-pixel imaging techniques. It uses Fourier basis patterns as structured patterns for spatial information acquisition in the Fourier domain. However, the spatial resolution of the image reconstructed by FSI mainly depends on the number of Fourier coefficients sampled. The reconstruction of a high-resolution image typically requires a number of Fourier coefficients to be sampled, and therefore takes a long data acquisition time. Here we propose a new sampling strategy for FSI. It allows FSI to reconstruct a clear and sharp image with a reduced number of measurements. The core of the proposed sampling strategy is to perform a variable density sampling in the Fourier space and, more importantly, the density with respect to the importance of Fourier coefficients is subject to a one-dimensional Gaussian function. Combined with compressive sensing, the proposed sampling strategy enables better reconstruction quality than conventional sampling strategies, especially when the sampling ratio is low. We experimentally demonstrate compressive FSI combined with the proposed sampling strategy is able to reconstruct a sharp and clear image of 256×256 pixels with a sampling ratio of 10%. The proposed method enables fast single-pixel imaging and provides a new approach for efficient spatial information acquisition.

## 1. Introduction

Single-pixel imaging [1-4] is a computational imaging technique which allows images to be acquired by using a spatially unresolvable detector, namely, single-pixel detector (such as, photodiode, solar cell, and photomultiplier tube). Compared with typical pixelated detectors (such as, CCD and CMOS), single-pixel detectors can work at a wide waveband, especially at the wavebands where pixelated detectors are expensive or even technically unavailable (such as inferred, deep ultraviolet, X-ray, or terahertz). Thus, single-pixel imaging has been considered as a potential solution for imaging at special wavebands and attracted a lot of attention in the last decade [5-12]. The core of single-pixel imaging is spatial light modulation. Spatial light modulation allows the spatial information of the target object to be encoded into a 1-D light signal sequence which is suitable for single-pixel detection. By decoding the spatial information from the resulting single-pixel measurements, the object image can be retrieved computationally.

Fourier single-pixel imaging (FSI) [13-18] is a branch of basis-scan single-pixel imaging techniques which use deterministic orthogonal basis patterns for spatial light modulation. In FSI, Fourier basis patterns (also known as sinusoidal intensity patterns) are used to acquire the Fourier spectrum of the object image. Compared with other basis-scan single-pixel imaging methods (such as, Hadamard [19-23], DCT [24]), FSI has been proven more data-efficient when the differential method of measurement is employed [25]. Specifically, differential FSI with 3-step phase shifting can reconstruct a noiseless image with single-pixel measurements which are 1.5 times the number of image pixels. Moreover, the generation of Fourier basis patterns is flexible. The basis patterns of FSI can be generated by the interference of two plane waves [26], which potentially allows FSI to be implemented without using a pixelated spatial light modulator. Such a property benefits imaging at the wavebands where spatial light modulators are not available.

However, as other single-pixel imaging methods do, FSI suffers from the tradeoff between imaging quality and imaging efficiency. The spatial resolution of the image reconstructed by FSI mainly depends on the number of Fourier coefficients sampled. Specifically, it requires more spatial information to reconstruct an image with finer details. And the more spatial information implies the more single-pixel measurements, and consequently, the longer data acquisition time. However, the data acquisition time is crucial for fast imaging, especially when imaging a dynamic scene. Thus, it is worth exploring how to improve the data efficiency in FSI.

Initially, FSI was proposed with a sampling strategy [13] which exploits the prior knowledge that the most information of natural images is concentrated in low-frequency bands of the Fourier space. According to the spiral sampling strategy, only low-frequency Fourier coefficients will be sampled with high-frequency coefficients discarded. But the lack of high-frequency components could result in severe ringing artifacts in the reconstructed images, especially when the sampling ratio is low. Later, several sampling strategies in the Fourier space have been proposed, such as statistical-importance [14], diamond [27] and circular [27]. Different sampling strategies are referred to different orderings of the Fourier basis patterns. We note that the research on basis patterns ordering is a hot spot in single-pixel imaging. For example, Russian doll [19], cake-cutting [20], origami [21], total variation ascending orderings [22] were recently proposed for Hadamard single-pixel imaging.

Given an ordering of patterns and a specific sampling ratio, only a sub-set of the complete set of basis patterns will be used for spatial light modulation. As a result, the reconstructed image is under-sampled. To improve the quality of under-sampled images in FSI, here we propose a Gaussian random sampling strategy. The core of the strategy is to perform a variable density sampling in the Fourier space and the density is based on the importance of Fourier coefficients. Specifically, the sampling density with respect to the importance of Fourier coefficients is subject to a 1-D Gaussian function. Here, the importance of a Fourier coefficient is evaluated by the magnitude of the modulus of the coefficient. In other words, the larger the modulus of a Fourier coefficient is, the more important this coefficient is. Combined with compressive sensing (CS), the un-sampled high-importance coefficients can be recovered by optimization. And more single-pixel measurement can be spent in the acquisition of the low-importance coefficients. In consequence, the spatial resolution of the reconstructed image is improved. The proposed method is able to reconstruct a clear and sharp image with a small number of single-pixel measurements relative to image pixel count. The proposed method enables fast single-pixel imaging and provides a new approach for efficient spatial information acquisition. We note that deep learning has been adopted into single-pixel imaging recently [28-35] for improvement of image reconstruction or reduction of data acquisition time. However, deep learning commonly requires a large amount of labeled data for training and the training process is generally computationally exhausted.

## 2. Principle

As Fig. 1 shows, FSI is based on the Fourier transform, using Fourier basis patterns for spatial light modulation. Fourier basis patterns are also known as sinusoidal intensity patterns, each of which can be expressed as

$$P(x,y) = \frac{1}{2} + \frac{1}{2} \cdot \cos\left[2\pi\left(f_x x + f_y y\right) + \varphi\right], \quad (1)$$

where $(x,y)$ denotes the coordinate in the spatial domain, $\varphi$ denotes the initial phase, $f_x$ and $f_y$ are spatial frequency at $x$ and $y$ direction, respectively. By employing the 3-step phase-shifting strategy, each Fourier coefficient, $\tilde{I}(f_x, f_y)$, can be acquired by using a set of 3 Fourier basis patterns of the same spatial frequency pair but a different initial phase. The initial phase, $\varphi_i$, of the $i$-th step pattern is $2(i-1)\pi/3$. The Fourier coefficient $\tilde{I}$ associated with the spatial frequency $(f_x, f_y)$ can be calculated through

$$\tilde{I}(f_x, f_y) = (2D_1 - D_2 - D_3) + \sqrt{3}\mathrm{j}(D_2 - D_3), \tag{2}$$

where $\mathrm{j}$ is the imaginary unit, and $D_i$ denotes the single-pixel measurement corresponding to the *i*-th step pattern. As Fig. 1(b) shows, the Fourier spectrum of a real-valued image is conjugate symmetric. Thus, the symmetric coefficients need not be sampled. To reconstruct a noiseless image by FSI, the number of Fourier coefficients acquired is one half of the number image pixels. If the 3-step phase shifting is employed for differential measurement, the number of single-pixel measurements will be 1.5 folds of the image pixels.

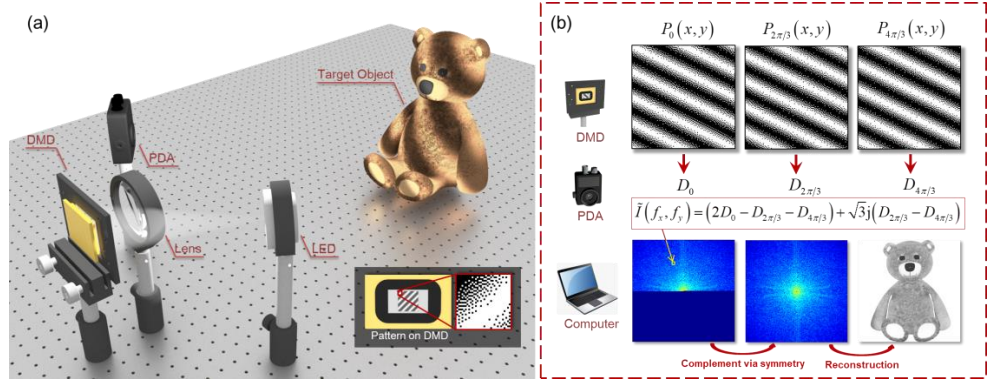

Fig. 1. Illustration of FSI with 3-step phase shifting. (a) In a structured illumination based setup, the target object is under illumination of Fourier basis patterns generated by a DMD. The Fourier basis patterns are dithered. DMD: digital micro-mirror device, PDA: photodiode amplified. (b) The target object image is retrieved by acquiring the Fourier spectrum of the image. Each complex-valued Fourier coefficient can be acquired by using a set of 3-step phase-shifting Fourier basis patterns. The phase shift is $2\pi/3$. The conjugate symmetry of the Fourier spectrum allows a noiseless image to be retrieved with only one half of the Fourier coefficients acquired.

However, the reconstruction of an image in a large size still needs a large number of single-pixel measurements. Undersampling is a widely used strategy to reconstruct an image of satisfactory quality with a reduced number of measurements. In the context of FSI, undersampling means only a portion of the Fourier coefficients are sampled. Here we propose a Gaussian random sampling strategy which allows the coefficients of higher importance to be sampled with a higher density while the coefficients of lower importance to be sampled with a lower density. Such a sampling strategy would result in a few high-importance coefficients being not sampled, but adopting a CS algorithm for image reconstruction allows those un-sampled but high-importance coefficients to be accurately recovered through optimization. It is because high-importance coefficients are sampled with a high density, which imposes a strong constraint to find the globally optimized solution for the un-sampled high-importance coefficients. As such, more single-pixel measurements can be spent in sampling the remaining low-importance coefficients and those low-importance coefficients generally contribute to high-frequency information. Consequently, the spatial resolution of the resulting image is improved.

However, it is difficult to predict which Fourier coefficients are important for any object or scene to be imaged. Here we adopt the method reported by L. Bian *et al*. [14] to derive the importance distribution of coefficients in the Fourier space for natural images. Specifically, we use the database DIV2K [36] for statistical importance analysis of Fourier coefficients. The database provides hundreds of high-resolution full-color natural images, and is typically used for convolutional network training in the computer vision community. As Fig. 2(a) shows, we use all 800 natural images provided in the training set of the database. Each high-resolution full-color image is converted into grayscale and segmented to a number of $M \times N$-pixel sub images, where $M$ and $N$ depend on the size of the reconstructed image. In our case, $M = 256$ and $N = 256$. The number of the resulting segmented sub images is 32,208. Then we apply a 2-D Fourier transform to every single sub image and sum up the moduli of all resulting Fourier spectra. Lastly, the Fourier coefficients of the summed Fourier spectrum are sorted in a descent order of magnitude. Please note that the conjugate symmetric coefficients are discarded. Thus, the number of the sorted coefficients is 32,770 in our case.

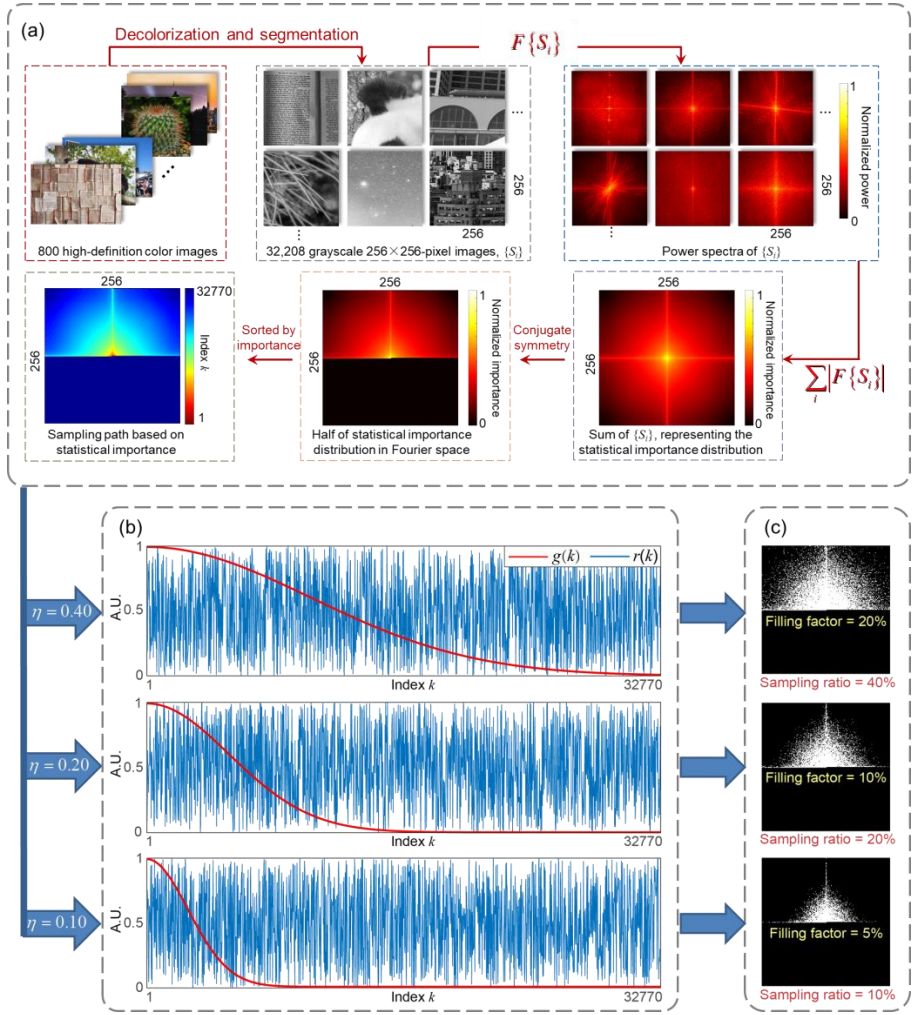

Fig. 2. The generation of Gaussian random sampling masks for different sampling ratios. In the first step (a), the descent order of the importance of Fourier coefficients is obtained via statistical analysis. Each coefficient has its own index $k$ which indicates the importance of the coefficient. The smaller index, the higher importance. For the second step (b), a uniformly distributed random function $r(k)$, and a Gaussian function $g(k)$ with a specific sampling ratio, $\eta$, are generated. For the third step (c), all $k$-th Fourier coefficients will be marked as 'to be sampled' (white pixel) in the sampling mask, if $g(k) > r(k)$. Filling factor is defined as the ratio of marked coefficients to all coefficients in the Fourier space, which is also one half of the sampling ratio (due to the conjugate symmetry).

Each sorted coefficient has its own index $k$ which indicates the importance of the coefficient. The smaller index is, the higher importance will be. Next, as Fig. 2(b) shows, we generate a uniformly distributed random function $r(k)$ whose range is from 0 to 1. We also generate a Gaussian function

$$g(k) = \exp\left\{-\left[(k-1)/k_{\max}\right]^2 / \sigma\right\}, \tag{3}$$

where $k$ is positive integer denoting the index of the descent importance order, $k_{\max}$ = 32,770 equal to the number of sorted Fourier coefficients in our case, and $\sigma$ is the standard deviation of the Gaussian function. The value of $\sigma$ depends on the sampling ratio $\eta$. Here, the sampling ratio is defined as twice the number of sampled Fourier coefficients to the number of total Fourier coefficients, where "twice" is for the conjugate symmetry. When $\eta < 0.5$, there is a simple relation between the standard deviation and the sampling ratio, that is, $\sigma = (2\eta)^2 / \pi$. As indicated by the red lines in Fig. 2(b), the 1-D Gaussian function is centered at $k = 1$. If $g(k) > r(k)$, then the $k$-th

Fourier coefficient is marked to be sampled, and vice versa. The resulting sampling masks for different sampling ratios are shown in Fig. 2(c). By acquiring the marked Fourier coefficients via 3-step phase shifting, a partially sampled Fourier spectrum of the object image can be derived. The final image is derived by adopting a CS algorithm for image reconstruction.

We note that such a sampling strategy would result in a few high-importance coefficients being not sampled, but adopting a CS algorithm for image reconstruction allows those un-sampled but high-importance coefficients to be accurately recovered through optimization. It is because high-importance coefficients are sampled with a high density, which imposes a strong constraint to find the globally optimized solution for the un-sampled high-importance coefficients. As such, more single-pixel measurements can be spent in sampling the remaining low-importance coefficients and those low-importance coefficients mainly contribute to high-frequency information. Consequently, the spatial resolution of the resulting image is improved.

## 2. Simulation

We validate the proposed method by numerical simulations. In the following simulations, we compare the proposed sampling strategy with other two representative strategies including radial sampling [37] and circular sampling [13]. In particular, the widely used conventional FSI is also taken into comparison. The conventional FSI employs the circular sampling strategy for data acquisition and an inversed 2-D Fourier transform (IFT2) for image reconstruction. The sampling masks generated by the sampling strategies for different sampling ratios are shown in Fig. 3. Please note that white pixels in the masks indicate the Fourier coefficients to be sampled. Except for the conventional FSI (circular + IFT2), L1-Magic [38] and IFT2 is respectively employed as the CS algorithm for image reconstruction. The parameters of L1-Magic are set to be default.

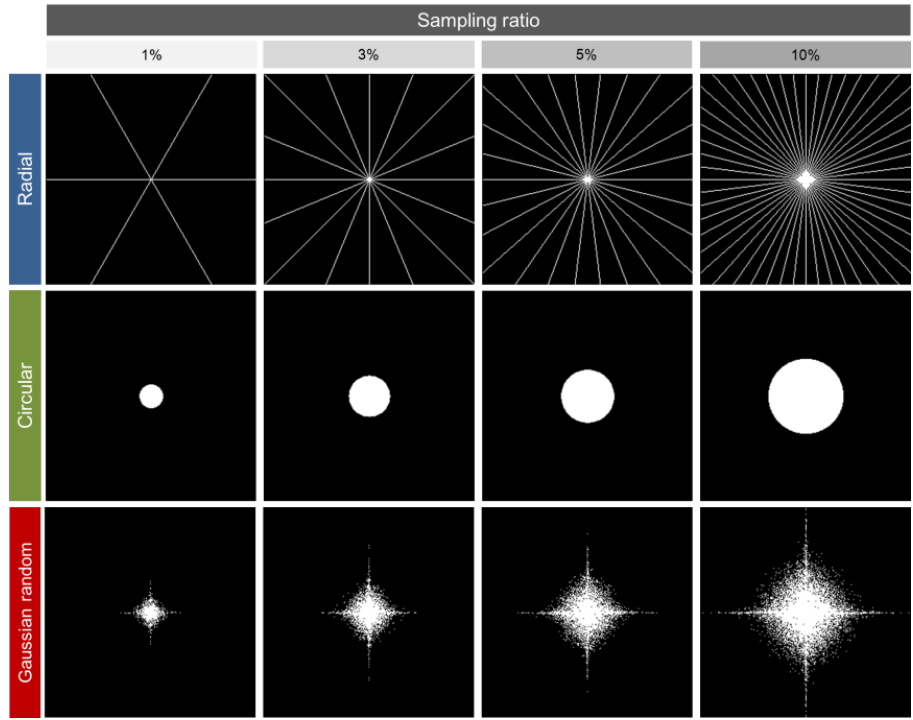

Fig. 3. Sampling masks used in the simulations and the experiments.

In the first simulation, a USAF-1951 resolution test chart is used as the target object. The target object image is with 256×256 pixels. As the results shown in Fig. 4, the radial sampling strategy is not able to reconstruct any bars, when the sampling ratio is below 10%. Even when the sampling ratio is 10%, the finest group that can be resolved is Group -2 Element 5. On the contrary, the circular sampling strategy and the proposed Gaussian random sampling strategy are able to give much better results. Specifically, the circular sampling strategy combined with IFT2 or CS can successfully reconstruct Group -2 Element 6, when the sampling ratio is 3%. The proposed sampling strategy combined with IFT2 or CS can even reconstruct Group -1 Element 1, but the result reconstructed by IFT2 is contaminated by noise. In other words, the spatial resolution enabled by the proposed method is 1.12 folds of those by the circular sampling based methods. We note that for a standard USAF-1951 resolution test chart the resolution (line pairs/millimeter) is defined as

$$resolution = 2^{\left(Group\ number + \frac{Element\ number - 1}{6}\right)}. \tag{4}$$

However, it is an enlarged USAF-1951 resolution test chart pattern that we use in the simulations. Thus, the resolution defined in Eq. (4) is used for the comparison of relative spatial resolution among different methods. As Fig. 4 shows, when the sampling ratio is 5%, both circular sampling based methods, and the Gaussian sampling strategy combined with IFT2 can only reconstruct Group -1 Element 2, while the proposed method can well reconstruct Group -1 Element 3. The resolution improvement is also 1.12 folds. When the sampling ratio is 10%, both circular sampling based methods, and the Gaussian sampling strategy combined with IFT2 can only reconstruct Group -1 Element 5, while the proposed method, as Fig. 5 shows, can well reconstruct Group 0 Element 2. The resolution improvement is 1.41 folds.

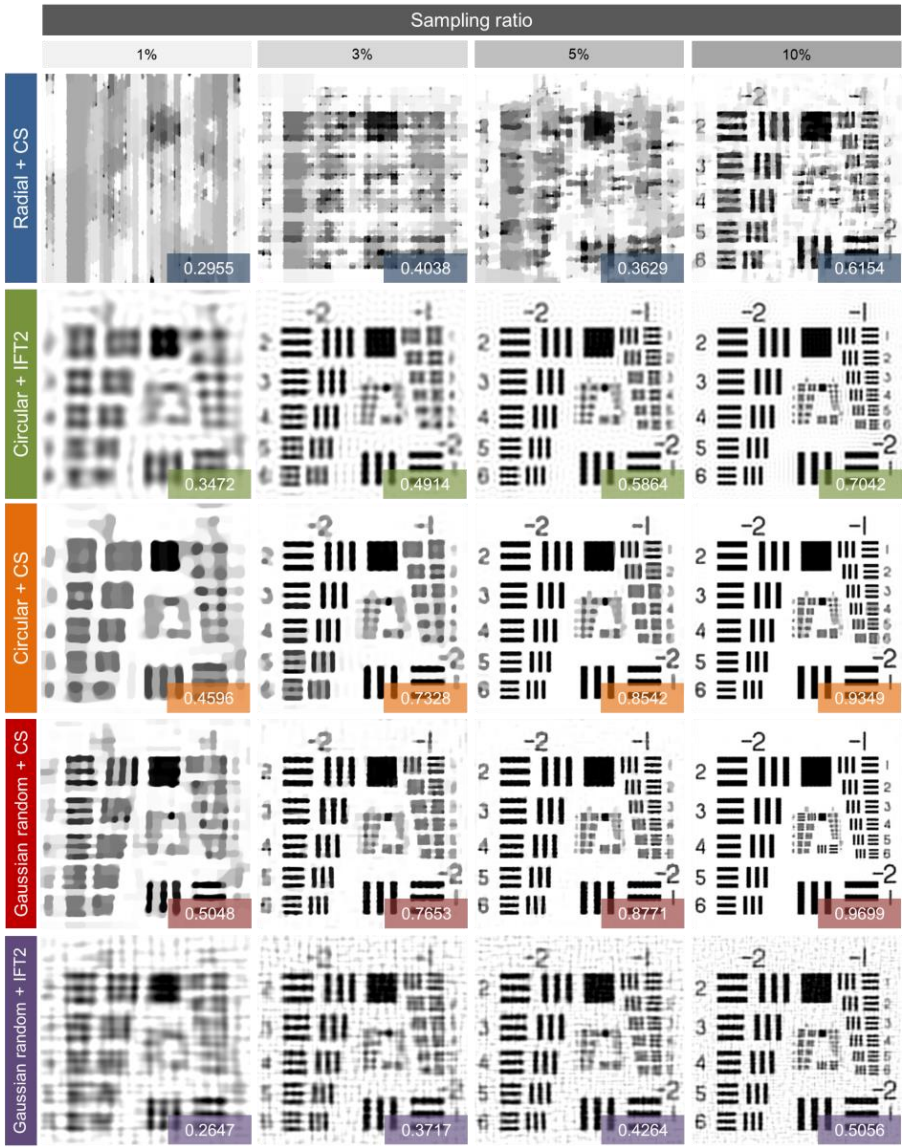

Fig. 4. Comparison of the reconstruction results of USAF-1951 resolution test chart for different sampling strategies and sampling ratios.

We also quantitatively evaluate the results by using structural similarity index (SSIM). The SSIMs also demonstrate the effectiveness of the proposed method. We therefore conclude that the proposed strategy enables higher spatial resolution than the other strategies in comparison. Both the visual quality and quantitative evaluation demonstrate that the proposed random sampling strategy can effectively improve the image resolution. The simulation also demonstrates that, for a given sampling strategy, CS can help in reducing ringing artifacts, but offers trivial improvement to image resolution.

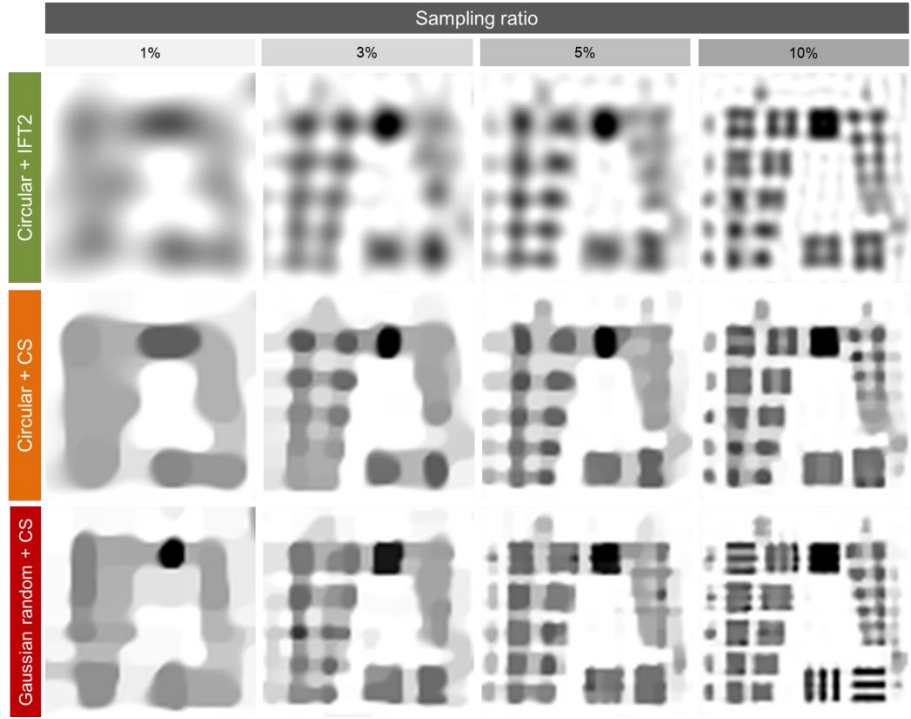

Fig. 5. Partial enlargements comparison for Groups 0 and 1 of the USAF-1951 resolution test chart for different sampling strategies and sampling ratios.

In the second simulation, we use "Cameraman" – a natural image for testing. The size of the test image is also 256×256 pixels. As the results shown in Fig. 6, the radial sampling strategy and the Gaussian random sampling strategy combined with IFT2 fail in reproducing a satisfactory result when the sampling ratio is below 10%. Even when the sampling ratio is 10%, the objects at the background are not able to be clearly reconstructed by both methods. In comparison with the circular sampling based method, the proposed method is able to reconstruct clear and sharp images. When the sampling ratio is 1%, although fine details are lost due to the ultra-low sampling ratio, the result by the proposed method looks sharper than the results by the circular sampling based method. When the sampling ratio reaches 3%, some details, the tripod and the face of the cameraman for example, have already been successfully reconstructed by the proposed method. When the sampling ratio reaches 5%, the tripod has been well reconstructed by the proposed method, while it looks blurred in the results by both circular sampling based methods. In particular, the result by the circular sampling strategy combined with IFT2 suffers from severe ringing artifacts. When the sampling ratio is 10%, the ringing artifacts can still be clearly seen in the image. On the contrary, no ringing artifacts can be seen in the images reconstructed by the circular sampling strategy combined with CS or the proposed method. As Fig. 6(b) shows, the result by the proposed method looks clearer and sharper.

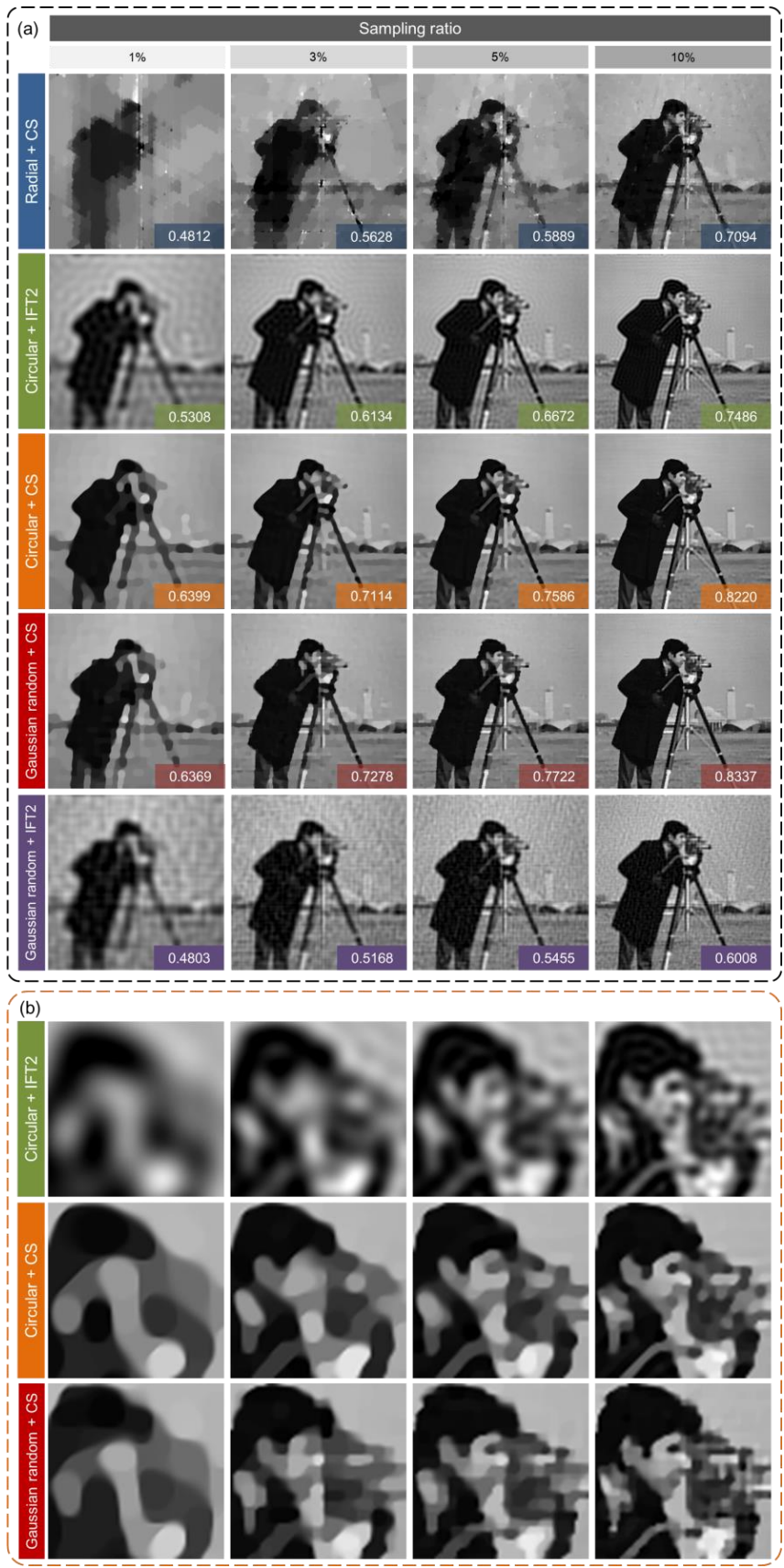

Fig. 6. Comparison of the reconstruction results of "Cameraman" for different sampling strategies and sampling ratios (a) and partial enlargements (b).

## 3. Experiment

We also demonstrate the proposed method with experiments. The schematic diagram of the experimental set-up is shown in Fig. 1(a). The set-up consists of a 12-watt white-light LED, a DMD (ViALUX V-7001), an imaging lens, a target object, a collecting lens, and a PDA (Thorlabs PDA101A). Note that we binarize the Fourier basis patterns with the upsample-and-dither strategy [25], so as to take the advantage of high-speed binary pattern generation offered by the DMD. The patterns are initially with 256×256 pixels. The patterns are upsampled with a ratio of 2 through the bicubic interpolation and then binarized using the Floyd-Steinberg algorithm. We use two different scenes for experiment. The one scene is a USAF-1951 resolution target printed on a piece of A4 paper. The other scene consists of a pair of china dolls as foreground and the printed resolution target pattern as background.

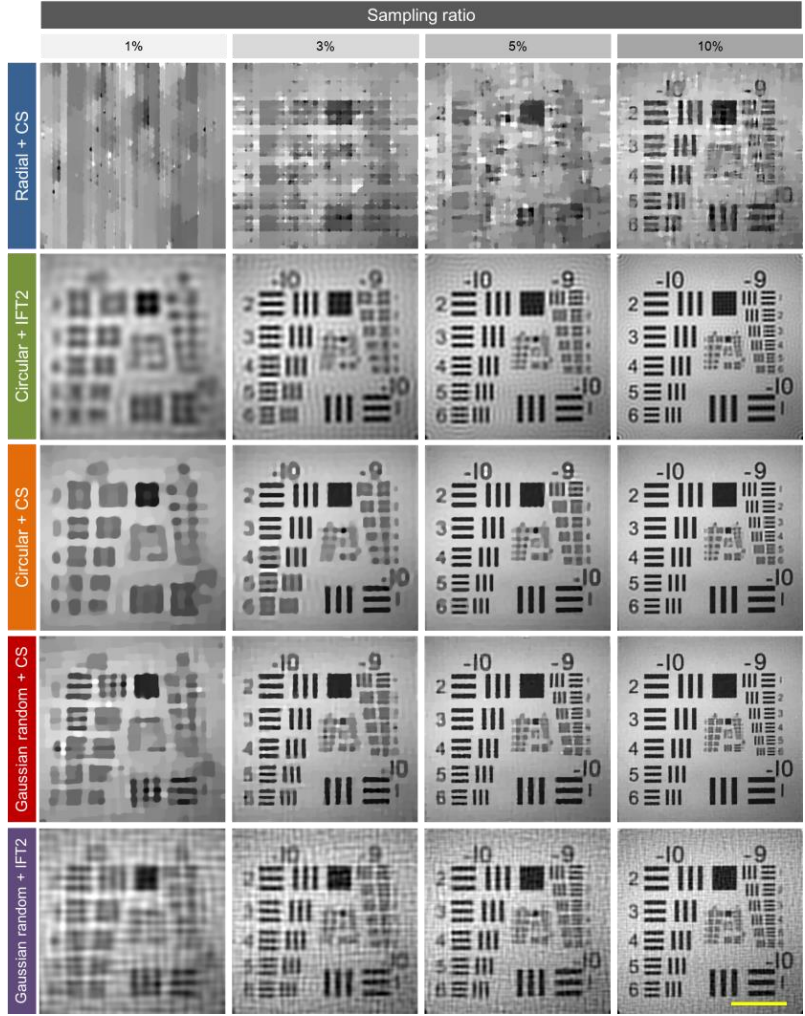

Fig. 7. Experiment results comparison for the scene of a printed USAF-1951 resolution test chart. DMD refreshing rate is 50 Hz. Scale bar = 5 cm.

Similarly, we compare the performance of the proposed method with other four representative methods in experiments. We note that it is an enlarged and printed USAF-1951 resolution test chart pattern that we use in the experiments. The scale bar of the pattern is given in Fig. 7. The reconstruction results of the two scenes are given in Figs. 7 and 8, respectively. As the figures show, the experiment results are consistent with those obtained in the simulations. For the resolution test chart scene, the proposed method is demonstrated able to offer higher spatial resolution than the circular sampling based methods and the Gaussian sampling strategy combined with IFT2. Specifically, the resolution improvements enabled by the proposed method are 1.26, 1.26, and 1.41 folds for the sampling ratio of 3%, 5%, and 10%, respectively.

For the china dolls scene, the images reconstructed by the proposed method look clearer and sharper than the results by the other four methods. We note that, in order to present a fair comparison of image resolution, the refreshing rate of the DMD is set to 50 Hz for a higher SNR in this experiment.

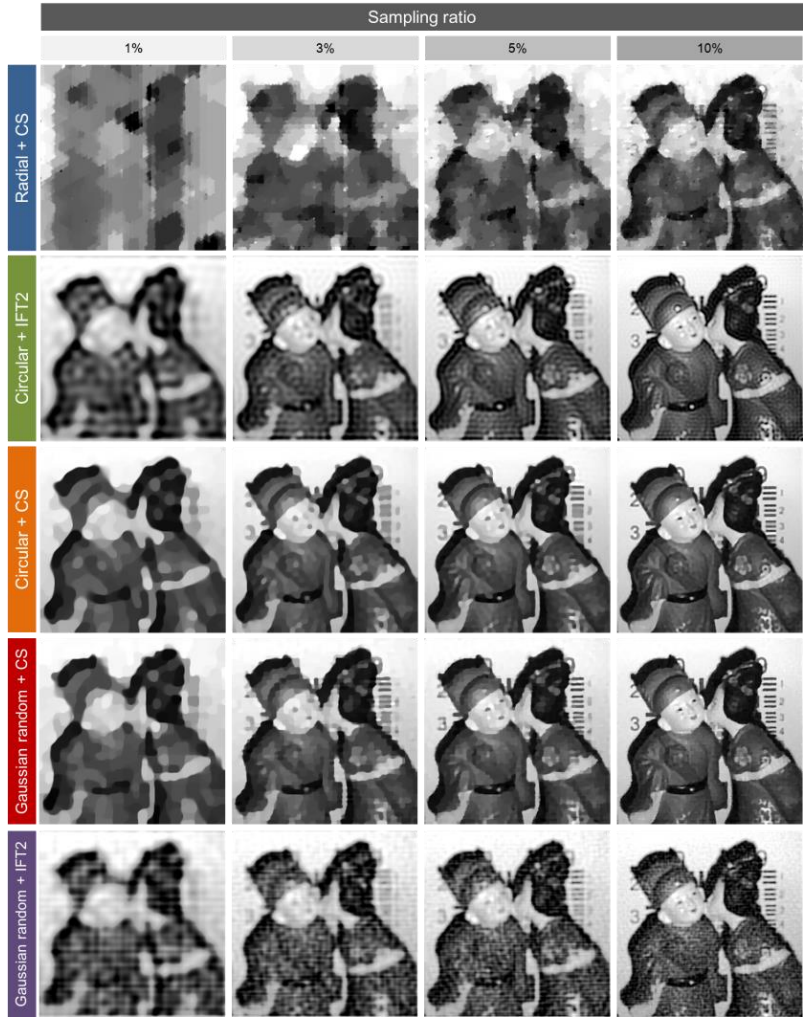

Fig. 8. Experiment results comparison for the scene of a pair of china dolls. DMD refreshing rate is 50 Hz.

To demonstrate the fast imaging ability of the proposed method, we test the method for different DMD refreshing rates. In this experiment, the sampling ratio is set to 10%, and therefore, the number of single-pixel measurements is 9,831. As the results shown in Fig. 9, the reconstructions for 50 Hz and 200 Hz are clear and without noticeable noise. In other words, the proposed method is able to capture a high-quality image of 256×256 pixels within 50 s. As the DMD refreshing rate increases, the noise becomes obvious and the signal-noise ratio (SNR) decreases. The image for 2,000 Hz is slightly noisy, but the data acquisition time can be reduced down to 5 s. When the DMD refreshing rate is 20,000 Hz, the image appears noisy, but the objects in the image are still recognizable.

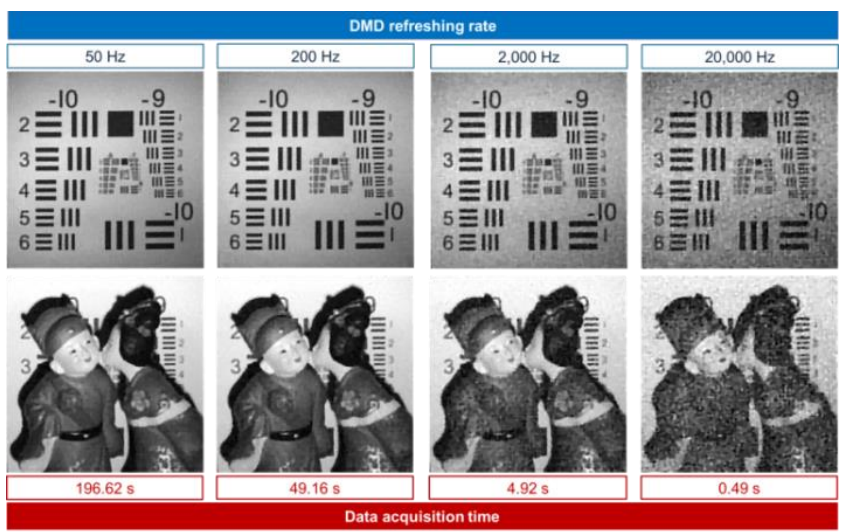

Fig. 9. Experiment results reconstructed by the proposed method for different DMD refreshing rates. Sampling ratio = 10%.

## 4. Discussion and Conclusions

Although the proposed method can reduce data acquisition time in comparison with conventional FSI, the image reconstruction time is increased due to the use of CS. We note that a short data acquisition time is more significant than a short image reconstruction time particularly in imaging a dynamic scene, because a short data acquisition time implies a higher temporal resolution, and therefore, less motion blurs. The proposed method might be able to accelerate by using an advanced CS algorithm.

Although the proposed method introduces randomness in data sampling, we conclude that the effect of randomness to the results is negligible. It can be evident by that we conduct the simulation by using the proposed method to reconstruct the "Cameraman" test image for 25 times with a sampling ratio of 10% and the standard deviation of the resulting SSIMs is only 0.0015.

We propose an efficient FSI method which enables high-quality image reconstruction with a reduced number of measurements. As demonstrated by the simulations and the experiments, the proposed method is able to reproduce a sharp and clear image of 256×256 pixels with a sampling ratio of 10%. This work benefits fast single-pixel imaging and provides a new approach for efficient spatial information acquisition.

## Funding

National Natural Science Foundation of China (NSFC) (61905098 and 61875074), Applied Basic Research Programs of Guangzhou (202002030319), and Natural Science Foundation of Guangdong Province (2018A030313912).

## Disclosures

The authors declare that there are no conflicts of interest related to this article.

## Competing interests

The authors declare no competing financial interests.